\documentstyle{l-aa}

\begin{document}

   \thesaurus{03         
              (13.10.1; 
               11.01.2: 
	       11.03.2  
               11.05.2  
               11.09.1  
               )} %

   \title{First detection of hotspot advance in a Compact Symmetric Object}

   \subtitle{Evidence for a class of very young extragalactic radio sources}

   \author{I.Owsianik 
          \inst{1}
          \and J.E.Conway
          \inst{2}
          }
            
  \offprints{I.Owsianik}

   \institute{ Toru\'n Centre for Astronomy,
                Radio Astronomy Dep.
                ul.Gagarina 11, PL87-100 Toru\'n, Poland\\
                e-mail iza@astro.uni.torun.pl
              \and 
                Onsala Space Observatory,
                S-439 92 Onsala, Sweden,\\
                e-mail: jconway@oso.chalmers.se\\
                 }

   \date{ Received 2 December, 1997; accepted 21 April, 1998}

   \maketitle
   
   \begin{abstract}

We present  the  results of  multi-epoch  global VLBI  observations  of the
Compact Symmetric Object (CSO), 0710+439 at 5 GHz.  Analysis of data spread
over 13 years shows strong  evidence for an increase in the  separation  of
the outer  components at a rate of $0.251 \pm 0.029 h^{-1}c$.  Dividing the
overall  size of  86.8  $h^{-1}$pc  by this  separation  rate  implies  an
estimated  kinematic  age of only  $1100 \pm 100$ yrs.  After  taking  into
account possible temporal variations in hotspot advance speeds due to cloud
collisions or hydrodynamic  instabilities  we argue that the upper limit to
the age of 0710+439 is most likely within a factor of $2$ of this  estimate
and certainly within a factor of $10$ (i.e.  $<11\,000$  yrs).  This result
therefore  strongly  supports the idea that Compact  Symmetric  Objects are
very young radio-loud sources.  Furthermore the large radiative  efficiency
we calculate  for 0710+439 is consistent  with strong  negative  luminosity
evolution as CSOs grow in size and with them evolving into classical double
sources.

   \end{abstract}


\section{Introduction}

Most  strong,  compact  ($<1\arcsec$)  radio  sources  when  imaged at high
resolution have a core-jet  morphology,  consisting of a bright  unresolved
core and a one-sided jet in which  superluminal  motion is often  observed.
These   core-jet   sources   are   thought  to  be  due  to  the  bases  of
relativistically  beamed  jets  orientated  close  to the  line  of  sight.
However  Phillips  \&  Mutel  (\cite{phillips})  first  identified  compact
objects which appeared to be dominated by two unbeamed emission  components
and  called  them  `Compact   Doubles'.  Conway  et  al.  (\cite{conway92})
described two similar  objects with more complex triple  structures.  Given
this  wider  range  of  morpholgies  Wilkinson  et  al.  (\cite{wilkinson})
renamed this whole class of radio  sources as `Compact  Symmetric  Objects'
(CSOs)  emphasising their primary property of symmetry.  Characteristically
these objects show high  luminosity  radio  emission  regions  separated by
$<1$kpc  which are  located  symmetrically  on both  sides of the centre of
activity.  It is thought  that  these high  brightness  regions  are due to
hotspots and minilobes  created by the  termination of oppositely  directed
jets and that this emission is free from relativistic beaming effects.  For
a recent  review of the  properties  of CSOs and related  sources see O'Dea
(\cite{odea98}).


 \begin{table*}[bt]
      \caption[]{Summary of  5 GHz global VLBI observations of 0710+439}
         \label{Tab.1}
        \begin{flushleft}
        \begin{tabular}{lllllllllllllll}
            \hline
            \noalign{\smallskip}
 Epoch &    1980.53 &   1982.93& 1986.89 &  1989.73 & 1993.44  \\
Duration (hr) &  11 & 12 &  4 $\times 1$  & 10&   4 $\times 0.6$ \\
Antennas$^{a}$ & BKGFo & BKGFo & SjBWKGFYo & SBWjKGFPkY & BLSWjNgROPHIY \\
Maximum baseline (M$\lambda$) & 136 &  136 &  136 &  138 & 146 \\
           \noalign{\smallskip}
            \hline
        \end{tabular}
        \end{flushleft}
\begin{list}{}{}
\item[$^{\rm a}$] S--- 26~m, Onsala Space Observatory, Onsala, Sweden;
j--- 26~\mbox{m},  MkII Telescope, Jodrell Bank, Cheshire, U.K.;
B---100~\mbox{m}, Max-Planck-Institute f\"ur Radioastronomie, 
        Effelsberg, Germany;
W---Westerbork Synthesis Radio Telescope, the Netherlands;
K---36.6~\mbox{m}, Haystack Observatory of the Northeast Radio Observatory 
          Corporation, Westford, MA;
G---42.7~\mbox{m}, National Radio Astronomy Observatory, Green Bank, WV; 
F---26~\mbox{m}, George R. Agassiz Station of Harvard University, Fort Davis, 
          TX;
Y---26~\mbox{m}, VLA, Socorro, NM;
o---40~\mbox{m}, Owens Valley Radio Observatory of the California Institute of 
          Technology, Big Pine, CA;
L---32~\mbox{m}, Istituto di Radioastronomia, Medicina, Italy;
N---32~\mbox{m}, Istituto di Radioastronomia, Noto, Italy;
g---22~\mbox{m}, Simeiz, Ukraine;
k---25~\mbox{m}, VLBA antenna, Kitt  Peak, AZ;
R---25~\mbox{m}, VLBA antenna, Brewster, WA;
O---25~\mbox{m}, VLBA antenna, Owens Valley, CA;
P---25~\mbox{m}, VLBA antenna, Pie Town, NM;
H---25~\mbox{m}, VLBA antenna, Hancock, NH;
I---25~\mbox{m}, VLBA antenna, N. Liberty, IA
\end{list}
  \end{table*}


From the  earliest  papers it was  suggested  that CSOs were young  sources
(Phillips  \&  Mutel  \cite{phillips}),  which  evolved  into  larger-sized
objects.  Alternatively  it has been  proposed  that  CSOs are  `frustrated'
sources, in which higher  density  and/or  turbulence  in the  interstellar
medium  inhibits  their  growth to larger  dimensions  (van  Breugel et al.
\cite{breugel}).  Finally  it has been  proposed  that they are a  separate
class of short lived  objects, which `fizzle out' after about  10$^{4}$ yrs
and do not grow to large sized objects (Readhead et al.  \cite{read94}).

Detailed  theories of the youth  model of compact  sources  show that it is
feasible that CSOs are part of an evolutionary sequence in which they later
evolve into the slightly larger Compact Steep Spectrum (CSS) sources, which
finally evolve into classical doubles (Fanti et al.  \cite{fanti}, Readhead
et al.  1996a,b).  De Young  (\cite{young}) and Begelman  (\cite{begelman})
have used simple  physical  models to confirm that CSO sources are probably
not  frustrated  and  confined  but  instead  evolving.  An obvious  way to
distinguish  between competing models is to try to measure or set limits on
the growth in overall size of CSOs and so  determine  their ages  directly,
which is the purpose of this paper.


\begin{figure*}[htbp]
        \vspace{11cm}
        \includegraphics{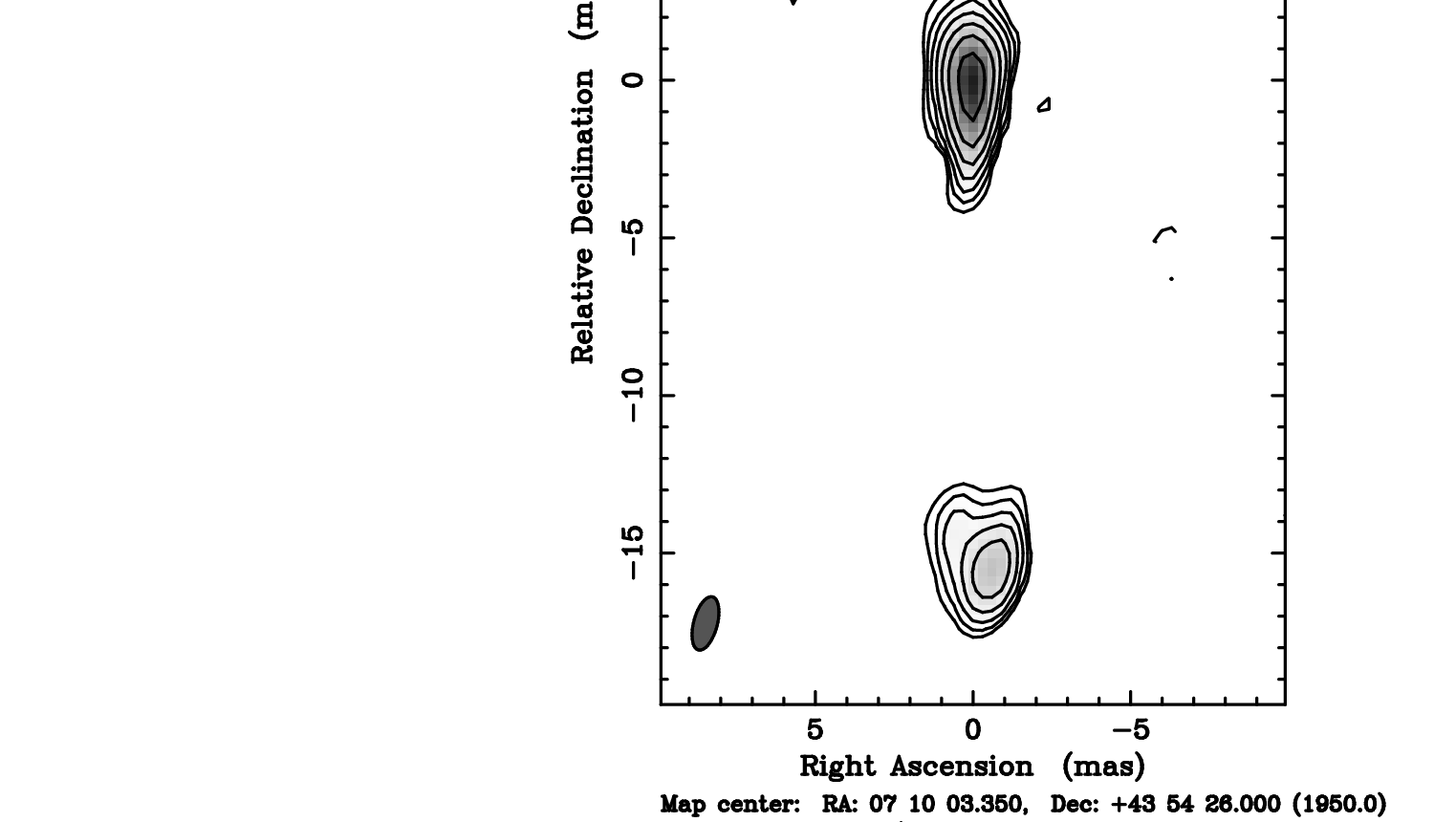}
        \includegraphics{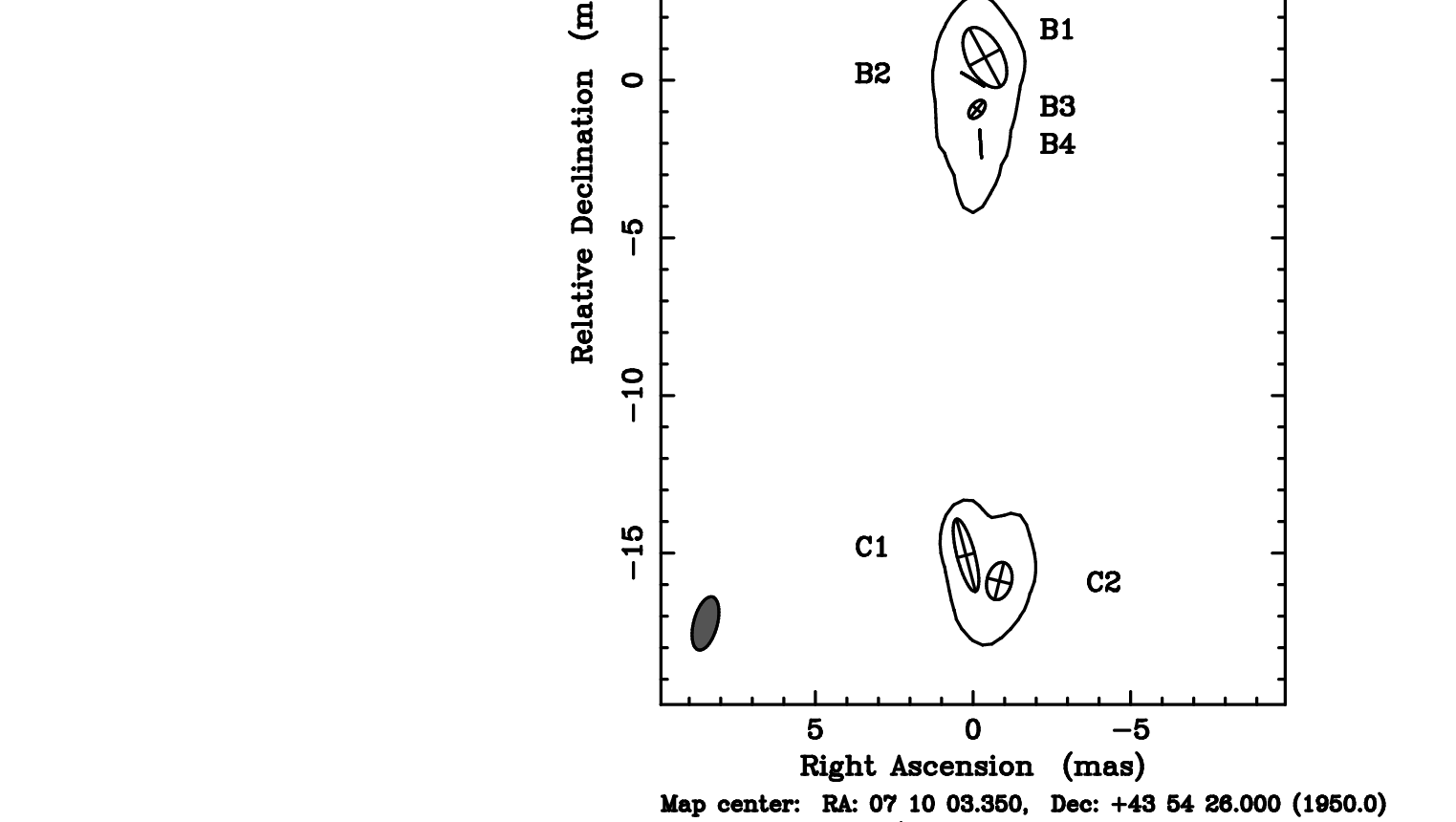}
        \vspace{5mm}
\caption{{\footnotesize {\bf a} Fifth epoch map of 0710+439. {\bf b} Diagram
  with positions and sizes of the Gaussian  model  fit  components}}
\label{Fig.1}
\end{figure*}

\section {General source properties}

\subsection {Optical properties}

The  radio   source   0710+439   has  been   identified   (Peacock  et  al.
\cite{peacock})  with a galaxy of {\it r} magnitude of $19.7 \pm 0.2$.  The
emission-line  redshift of the galaxy is {\it  $z=0.518$ } (Lawrence et al.
\cite{lawrence}).  At  this  redshift  1 mas  = 3.6  $h^{-1}$  pc  assuming
$H_{\circ}$ = 100 $h$ km s$^{-1}$  Mpc$^{-1}$  and  q$_{\circ}$ = 0.5.  The
optical spectrum (Lawrence et al.  \cite{lawrence})  shows absorption lines
characteristic  of an evolved stellar  population and an optical  continuum
shape typical of an elliptical galaxy without any evidence for a nonstellar
component.

\subsection {Radio properties}

0710+439 has high radio luminosity  (L$_{\mbox{5GHz}}$  = $5{\times}10^{33}
h^{-2}$ erg s$^{-1}$  Hz$^{-1}$;  Wilkinson et al.  \cite{wilkinson}).  The
flux density is very weakly  polarised ($< 0.15\% \pm 0.11\%$ at 5 GHz) and
the  observed  variations  of the flux  are not  statistically  significant
(Aller et al.  \cite{aller}).  The total  angular  size of the source is 29
mas, which  corresponds  to a projected  linear size of 86.8  $h^{-1}$ pc.
The source has been  mapped by VLBI at  several  frequencies  i.e.  1.6, 5,
10.7, and 15 GHz.  These  observations  showed the overall triple structure
of  the  source.  Despite   having   three   components   this  source  was
provisionally  classified  as a compact  double based on the fact that more
than 80\% of the emission came from two almost  equally  bright  components
(Pearson  \&  Readhead  \cite{pear88}).  Conway  et  al.  (\cite{conway92})
argued that the two outer components were hotspots and minilobes, while the
centre of activity was associated with the middle  component,  based on its
compactness,   spectrum   and  weak   flux   density   variability.  Recent
multi-frequency   observations  (Taylor  et  al.  \cite{taylor})  reveal  a
compact  component with a strongly inverted spectrum at the southern end of
the middle  component,  suggesting  that the true centre of  activity  lies
there.

\section {Observations and imaging}

\subsection {Observations}

While at most frequencies  0710+439 has only been observed once, at 5 GHz it
has been observed with a global VLBI array at 5 epochs spread fairly evenly
over a period of 13 years (see  Table~\ref{Tab.1}).  The first three epochs
were analysed by Conway et al.  (\cite{conway92}).  Here we reanalyse these
first three epochs and add new data from two additional epochs; a global 10
station  long  track   observation  made  on  25th  September  1989  and  a
multi-snapshot 13 station global observation made on 11th June 1993.

\subsection {Data reduction}

Observations  in all epochs were made in left circular  polarisation  (IEEE
convention)  and a bandwidth of 2 MHz was recorded using the MkII recording
system  (Clark  \cite{clark}).  The data  were  cross-correlated  with  the
JPL-Caltech VLBI Processor.  The data were fringe-fitted in AIPS (Schwab \&
Cotton \cite{schwab}) and averaged to 1 minute; error bars for the averaged
data  were  estimated  from  the  internal  scatter  of the data  over  the
averaging  interval.  Amplitude  calibration  for each  antenna was derived
from  measurements  of the antenna gain and system  temperature  during the
observations.

After  amplitude  calibration  the data were edited and mapped  using IMAGR
(AIPS package 1995) and Difmap  (Shepherd et al.  \cite{shepherd95}).  Many
iterations  of  phase   self-calibration  were  performed  before  applying
amplitude  self-calibration  at the end.  Windows for clean components were
added to provide  support and reject  sidelobes.  Initially  each epoch was
mapped separately starting with a point source model.  The fitted restoring
beams at each epoch were  typically  1 mas in the  East-West  direction  and
1.5 mas in the North-South  direction.  The highest  dynamic range image was
obtained  from the 5th epoch data set (see  Fig.~\ref{Fig.1}a).  Given this
best map we  therefore  remade maps at all epochs  using it as the starting
model for  self-calibration.  As described in a Sect.  4 we then used these
images to detect or set limits on internal motions within 0710+439.

Modelfitting  with  gaussian   components  was  also  carried  out  to  the
visibility  data at each epoch using the program  MODELFIT  in the  Caltech
VLBI package  (Pearson  \cite{pear91}),  which fits to the  amplitudes  and
closure  phases  directly, and also with the gaussian  modelfitting  option
within  Difmap.  The latter  fits  amplitudes  and phases but allows  phase
self-calibration  against the model, so that the model can  converge to fit
the closure  phases.  In all cases it was  possible to obtain  good fits to
the data using only a few Gaussian  components.  The  modelfitting  process
was started  using  gaussian  components  fitted to the 5th epoch CLEAN map
using the AIPS task JMFIT.  After varying all the gaussian  parameters  the
best  fitting  model to the fifth epoch data  contained 9  components  (see
Fig.~\ref{Fig.1}b, Table~\ref{Tab.2}).  This model provided a good fit with
reduced Chi-squared agreement factor for amplitude  Q$_{AMP}$=1.258 and for
closure-phases  Q$_{CLP}$=1.180  (for  definition  of agreement  factor see
Henstock et al.  \cite{henstock}).  To characterise temporal changes in the
source we  obtained  fits to the data at each  epoch,  using this 5th epoch
model as a starting point (see Sect.  4.1).


 \begin{table*}
     \begin{center}
      \caption[]{Model for the 5th epoch of 0710+439}
         \label{Tab.2}
         \begin{flushleft}
         \begin{tabular}{llllrrrrrrrrrrrrrrrrrrrrrrrllrrrrr}
            \hline
            \noalign{\smallskip}
& Component & & & S & & & & r & & & & $\Theta$ & & & & a & & & & $b/a$ & & &  & $\Phi$ \\
& & & & (Jy) & & & & (mas) & & & & ($\degr$) & & & & (mas) & & & & & & & & ($\degr$) \\
            \noalign{\smallskip}
            \hline
            \noalign{\smallskip}
& A1... & & & 0.069 & & & & 6.654 & & & & 6.261 & & & & 1.439 & & & & 0.556 & & & & -13.980 \\         
& A2... & & & 0.478 & & & & 8.287 & & & & -1.670 & & & & 0.840 & & & & 0.675 & & &  & 77.764 \\  
& A3... & & & 0.251 & & & & 7.623 & & & & -9.046 & & & & 2.472 & & & & 0.504 & & & & 120.877 \\

& B1... & & & 0.093 & & & & 0.814 & & & & -27.553 & & & & 2.081 & & & & 0.537 & & & & 28.275 \\
& B2... & & & 0.274 & & & & 0.028 & & & & 15.555  & & & & 0.831 & & & & 0.003  & & & & 59.254 \\
& B3... & & & 0.214 & & & & 0.932 & & & & -172.195 & & & & 0.669 & & & & 0.595 & & & & -39.600 \\
& B4... & & & 0.144 & & & & 2.033 & & & & -173.202 & & & & 0.878 & & & & 0.000 & & & & 3.603 \\

& C1... & & & 0.049 & & & & 15.070 & & & & 179.154  & & & & 2.370 & & & & 0.239 & & & & 14.660 \\
& C2... & & & 0.153 & & & & 15.907 & & & & -176.993 & & & & 1.210 & & & & 0.645 & & & & 166.070 \\
         \noalign{\smallskip}
         \hline
        \end{tabular}
        \end{flushleft}
\end{center}
\begin{list}{}{}
\item[] 
Parameters of the Gaussian components: S---flux density;
r,$\Theta$---polar coordinates, with polar angle measured from the North
through East;
a,b---major and minor axes of the FWHM contour;
$\Phi$---position angle of the major axis measured from the North through East
\end{list}
\end{table*}


\subsection {Source structure}

The CLEAN map of the 5th epoch data (rms noise = 0.9 mJy beam$^{-1}$) shows
clearly the overall triple structure of the source (see Fig.~\ref{Fig.1}a).
Maps at each epoch show three main components  which we name (from North to
South) A, B and C.  Each of these main components shows substructure  which
is represented in the modelfits as separate gaussians (e.g.  A1, A2 and A3;
see Fig.~\ref{Fig.1}b).

The CLEAN maps show that the northern   (A) and the southern  (C)
components show some faint extended  emission  around them.  In addition to
this both  modelfitting and imaging  indicate a compact feature (A2) within
component  A which we  interpret  (see Sect.  5.1) as a hotspot.  The CLEAN
images suggest a weak bridge of emission between A and the middle component
B, however imaging  simulations  show that this feature may not be reliable
(see  Appendix).  We were not able to  detect a  similar  jet-like  feature
connecting the middle and southern components, which was found on a 1.6 GHz
map (Xu \cite{xu}), possibly due to lack of surface brightness sensitivity.
We were also not able to detect any  emission  located to the East of the C
component,   which  is  seen  on  the  maps  made  by   Wilkinson   et  al.
(\cite{wilkinson}).

\begin{figure*}[htbp]
        \vspace{8.8cm}
        \includegraphics{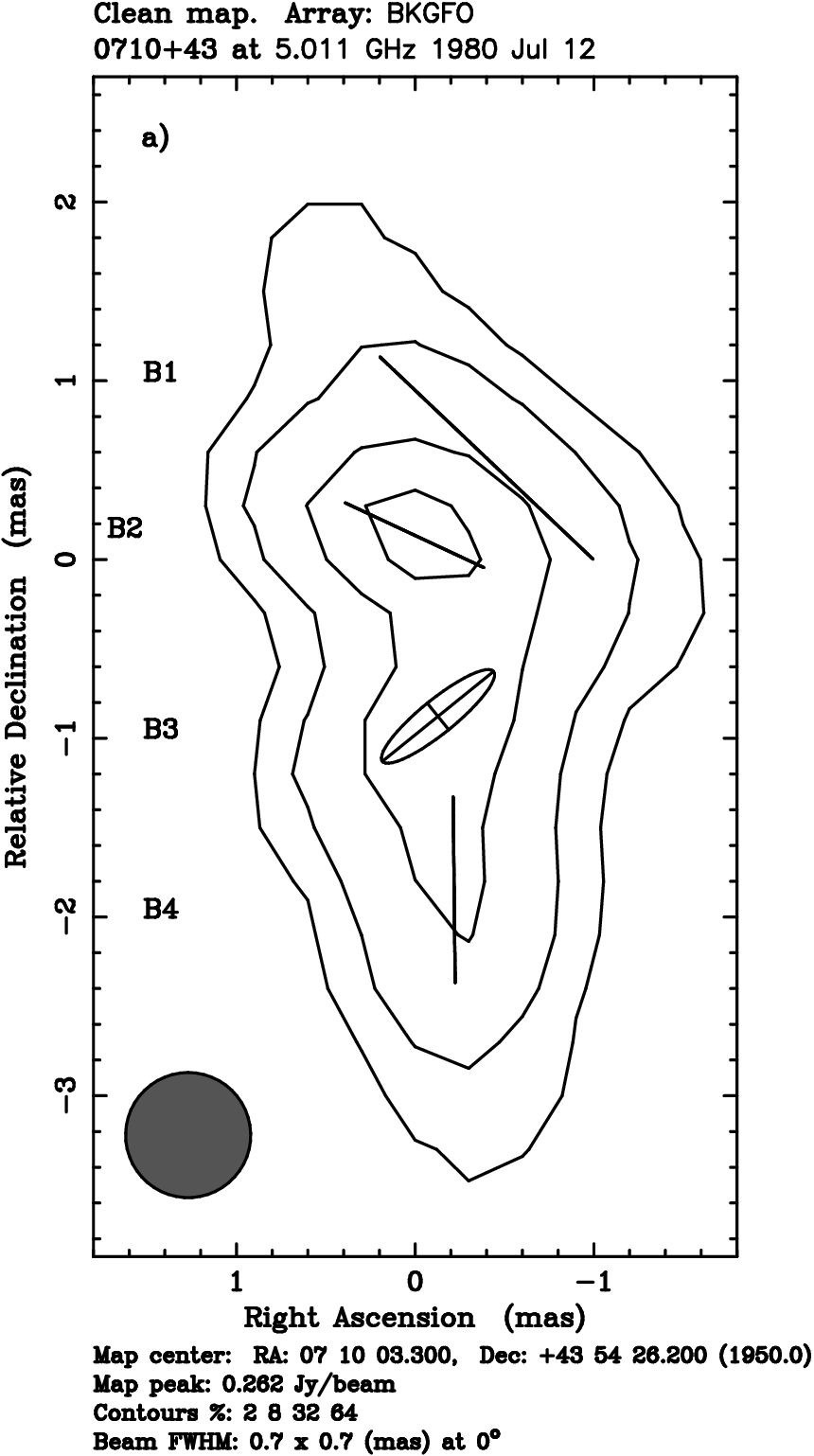}
        \includegraphics{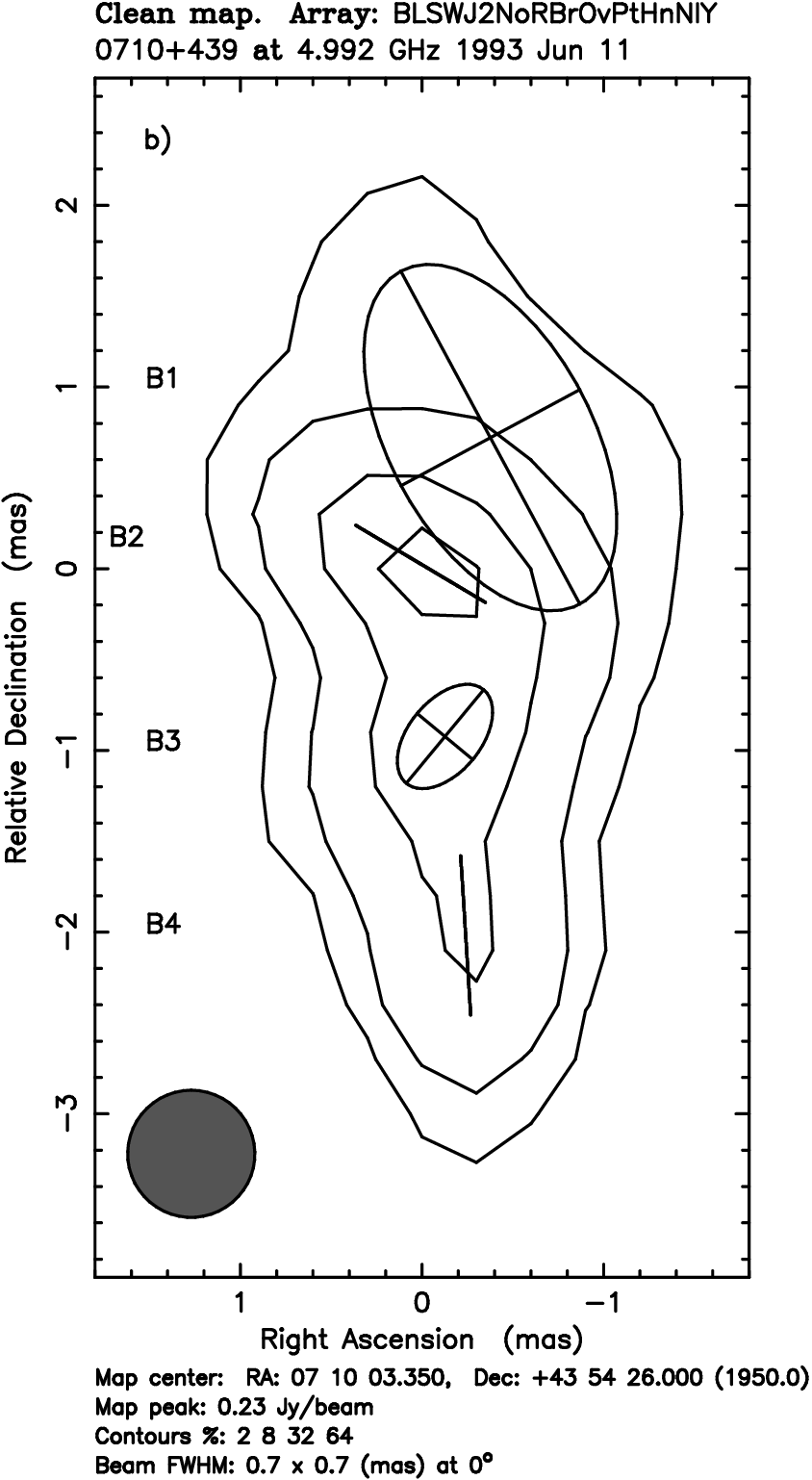}
        \vspace{5mm}
        \caption{{\footnotesize {\bf a and b.} Modelfit gaussians within middle
        component of 0710+439 at 5 GHz convolved with a 0.7 mas restoring beam:
        {\bf a} First epoch, {\bf b} Fifth epoch
                }}
        \label{Fig.2}
\end{figure*}

Fitting   the   B    component    required   4   gaussian    subcomponents.
Figure.~\ref{Fig.2}a,b  show these  gaussians as fitted to the 1st and 5th
epochs  respectively  (convolved with a circular restoring beam of FWHM 0.7
mas).  These  images show that the B  component  is narrow at the South and
becomes  wider to the North.  There also appears to be a slight kink in the
jet with the major axis of the B3 component  inclined at $45^{\circ}$  with
respect to the B2-B4  direction.  A similar kink is seen in the 15 GHz maps
and  models  (Taylor et al.  \cite{taylor}).  Despite  being  separated  by
almost 13 years the model components for 5 epochs  are very similar.
One possible  change is in the size of B1, however we conclude based on our
imaging  simulations that this may be an imaging  artifact.  In Sect.  4 we
describe  our  detailed  analysis  of  the  multi-epoch  data  which  shows
components  B2 and B4 to be  stationary  relative  to  each  other,  with a
possible small northward motion of B3.

Overall our maps and  modelfits to B agree closely with those  estimated at
15 GHz by Taylor et al.  (\cite{taylor}).  However in our  modelfits we did
not require a compact component at the position of the core seen at 15 GHz.
This is not unexpected  given that at high  frequency this  component has a
self-absorbed spectral index of  $\alpha_{15GHz}^{8.4GHz}$  = $1.6 \pm 0.4$
(Taylor   et   al.  \cite{taylor}),   where   flux   density   S   $\propto
{\nu}^{\alpha}$.  At 4.9 GHz with this  spectral  index we would expect the
core to have a flux  density  of  between  4.8 mJy  and  11.8  mJy.  At the
position  of the 15 GHz core on a 5th  epoch  super-resolved  map  (0.5 mas
FWHM, not shown) we did see a component  with flux density $4.3 \pm 1$ mJy,
if this feature is real it implies a spectral index of $2.1\pm 0.2$ between
4.9 GHz and 15 GHz  consistent  with a synchrotron  self-absorbed  spectrum
core component.


\section {Multi-epoch intercomparison}

\subsection {Procedure}

A serious problem with analysing data from VLBI  observations is that there
is significant freedom in making images.  Data are degraded by instrumental
errors,  incomplete and different  aperture  coverages, and  ambiguities in
deconvolution  and  self-calibration,  all of which can strongly affect the
final  results.  For this  reason  intercomparison  of models and maps made
separately at each epoch is not a good method of detecting small changes in
a source (Conway et al.  \cite{conway92}).

To  minimise  the above  effects we used the 5th epoch  model and CLEAN map
(see   Fig.~\ref{Fig.1}a,b)  as  starting  points  in  re-modelfitting  and
re-mapping  all of the epochs.  This method  should  limit the  differences
between the final  images of all five epochs of 0710+439, so we can be sure
that any  differences  seen are  demanded  by the  data and are due to real
changes in the structure of the source (see Appendix).

The detailed procedure in modelfitting at each epoch was, starting with the
5th epoch model, to first allow just the flux density of all  components to
vary; however in each case the fit remained poor.  We next allowed  changes
of all  parameters  of the  gaussians  within the B component  (e.g.  flux,
radius, $\Theta$, major axis, axis ratio, and $\Phi$) which gave a somewhat
better fit, but only after  allowing all the components to move in position
(which led to significant  motion mainly in component A2) did we get a good
fit.  Finally the {\it u-v} data were amplitude self-calibrated against the
model  and  one  final  iteration  of  modelfitting  carried  out in  which
component  positions were again allowed to vary.  We note that it was never
necessary at any epoch to change the size or shape of the gaussians  within
the A or C components.  The final models had good agreement  factors to the
data  (for  epochs 1 to 5 total  agreement  factors  were  Q$_{TOT}$=1.149,
Q$_{TOT}$=1.047,   Q$_{TOT}$=1.039,   Q$_{TOT}$=1.145  and  Q$_{TOT}$=1.224
respectively).

\subsection {Component position and flux variations}

\begin{figure}[htbp]
        \vspace{8cm}
        \includegraphics{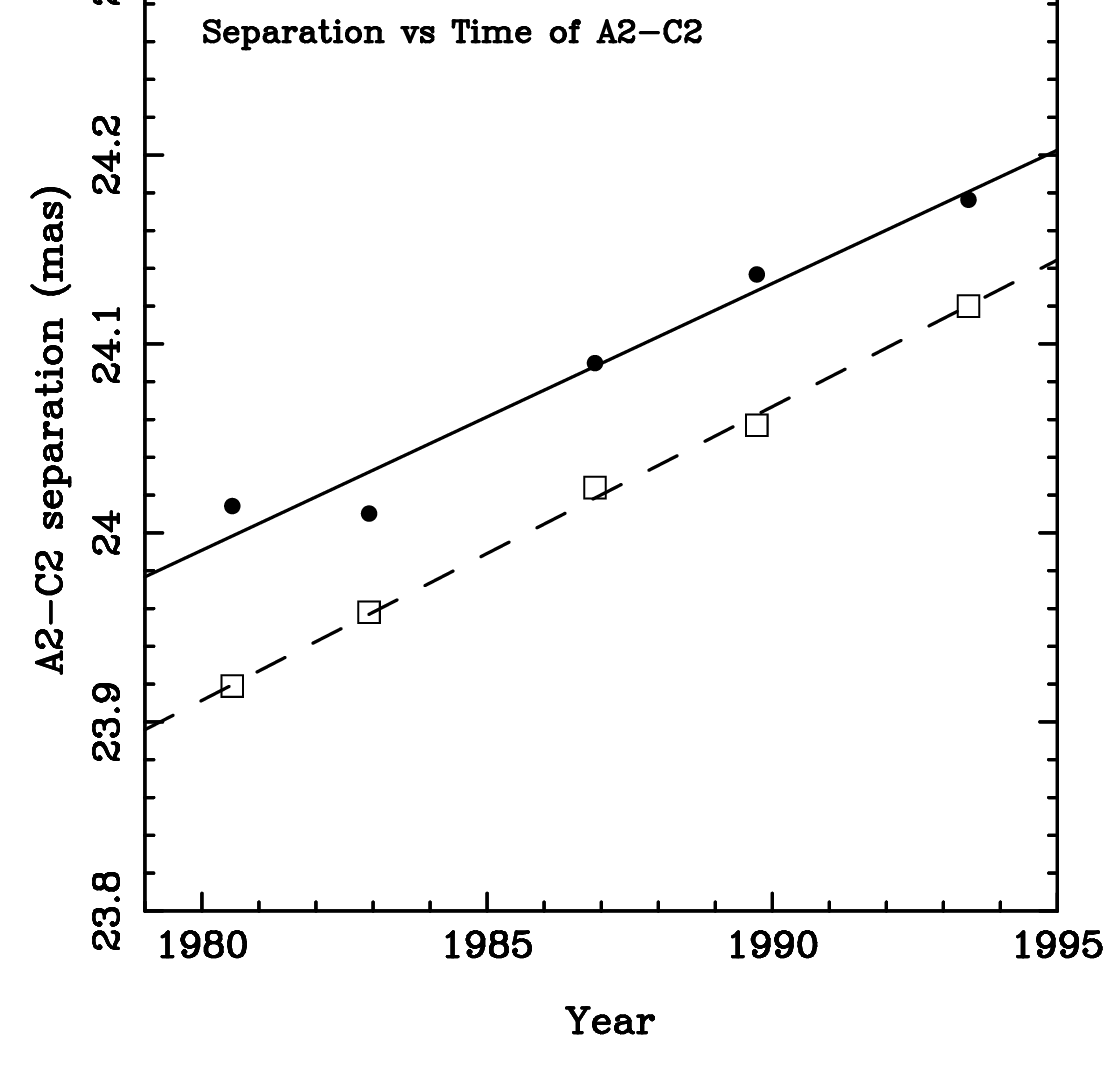}
        \vspace{5mm}
\caption{{\footnotesize Changes in separation with time   
                of the components A2 and C2. Filled circles represent data 
                obtained by MODELFIT, solid line shows linear regression fit 
                to these data. Open squares
                represent data obtained by JMFIT, dashed line shows linear 
                regression fit. See Sect. 4.2 for a discussion of the
                errors on the plotted points and the fits}}
\label{Fig.3}
\end{figure}

From the  modelfits we measured the  separation of many pairs of components
as a function of time and fitted  linear  regression  lines to this data to
estimate  relative  velocities  (see  Table~\ref{Tab.3}, 
Fig.~\ref{Fig.3}).

Although in VLBI data analysis  various attempts have been made to estimate
{\it a priori}  error bars on  component  positions  these  schemes  are of
doubtful  reliability.  Gaussian error bars estimated from the  variability
of reduced  Chi-squared on moving the components  critically  depend on the
number of  degrees  of  freedom  in the data which  depends  in turn on the
unknown degree of correlation of phase and amplitude  errors with time.  In
addition as noted in Sect.  4.1 even our best  fitting  model has a reduced
Chi-squared  which is much further from unity than would be expected  given
statistical  arguments.  Given this  situation we chose instead to estimate
errors on motions from the internal  scatter of our separation  versus time
data using standard methods of linear  regression  analysis.  These methods
applied to our data suggest  that for the  brighter  components  the random
errors on the  relative  separation  in each epoch are of order  20$\mu$as.
This is comparable to the  estimates we obtained  from imaging  simulations
(see Appendix A.2).


We find from our fitting  that there is no  evidence  for  relative  motion
between any of the components  A1, A3, B1, B2, B4, C1 and C2 (we will refer
to these  components  as the  `stationary'  group).  In  contrast  the most
significant  separation we find is between the outer  components A2 and C2.
From linear regression  analysis of MODELFIT data we find a separation rate
between these two components of $14.116 \pm 1.606$  $\mu$as  yr$^{-1}$ (see
Fig.~\ref{Fig.3}).  These components are well separated on our CLEAN images
which  allows us to also use the AIPS task JMFIT to fit the  position of A2
and C2 on the CLEAN images at each epoch, giving a similar  separation rate
of $15.538 \pm 0.445$ $\mu$as yr$^{-1}$.

Given the small number of degrees of freedom tests of significance are best
made from examining the correlation  coefficients  obtained from our linear
regression  analysis.  For the  MODELFIT  and JMFIT  analysis  of the A2-C2
separation we obtain correlation  coefficients between epoch and separation
of 0.981  and  0.998  respectively,  which  allows  us to  reject  the null
hypothesis of no motion at better than the 1\% and 0.1\% confidence levels.
Analysis of other pairs of components (see Table~\ref{Tab.3}) suggests that
the significant change in A2-C2 separation is caused primarily by motion of
A2 northward rather than motion of C2 southward.  For instance we found the
separation rate of A2-B2 to be $12.913 \pm 2.555$ $\mu$as  yr$^{-1}$,  very
similar to the A2-C2  separation  rate.  In contrast  the B2-C2  separation
rate of $1.370 \pm 3.383$ $\mu$as yr$^{-1}$ is consistent with zero.

Amongst the other components the only other indication of motion is that B3
is  moving  northward  relative  to B2 (and  B4 and  other  members  of the
stationary  group) at a rate of $6.089 \pm  1.752$  $\mu$as  yr$^{-1}$.  We
also   searched  for  motions   between   component   pairs  in  directions
perpendicular  to the  vectors  separating  them but  found no  significant
motions.

 \begin{table}
      \caption[]{Measured apparent motions of the gaussian components within 
      0710+439}
         \label{Tab.3}
        \begin{flushleft}
        \begin{tabular}{llrlll}
            \hline
            \noalign{\smallskip}
           & Components & Apparent & &  & Estimated \\
           &     &  velocity & & & $1\sigma$ error \\
           &     &   [$h^{-1}c$] & &  &  [$h^{-1}c$] \\
          \noalign{\smallskip}
          \hline
& A2-A1  & 0.178  & & & 0.217 \\
& A2-A3  & 0.167  & & & 0.057 \\
& A2-B2  & 0.230  & & & 0.053 \\
& A2-B4  & 0.258  & & & 0.112 \\
& A2-C2  & 0.251  & & & 0.029 \\
          \noalign{\smallskip}
          \hline
	\noalign{\smallskip}
& B2-B1  & 0.194   & & & 0.160 \\
& B2-B3  & -0.107 & & & 0.032 \\
& B2-B4  & 0.032  & & & 0.068 \\
& B2-C2  & 0.025  & & & 0.061 \\
          \noalign{\smallskip}
          \hline
          \noalign{\smallskip}
& C1-C2 &  -0.353  & & & 0.253 \\
            \hline
        \end{tabular}
        \end{flushleft}
\end{table}


Finally in our analysis of the multi-epoch  data we searched for variations
in component  flux  densities.  In order to eliminate the effects of errors
on the overall  flux density  scale at each epoch we measured  the ratio of
each  component's  flux  density  to  that of  component  C2.  None  of the
`stationary' components showed significant flux variations relative to each
other or to C2, strongly arguing that all these components  stayed constant
in flux density over the observing  period.  We did however  detect  strong
variability  of the flux ratio for A2/C2 (see  Fig.~\ref{Fig.4}),  implying
changes in the A2 flux density and also a possible  steady  increase in the
flux  density of B3 of about $24\% \pm 8\%$  between the first and the last
epochs.

\subsection {Bulk motion or internal structure changes?}

The change in separation of the outer  components of 0710+439 over 13 years
is  approximately  $1/7$ of the  beam  FWHM in the  North-South  direction.
However we note that this shift is 1/3 of the FWHM of the A2  component  in
the same  direction.  This  large  shift  combined  with the  fact  that it
appears to be consistent from epoch to epoch (see  Fig.~\ref{Fig.3}) and is
the same when measured  relative to several gaussian  components,  strongly
argues that the motion of component A2 is real.

\begin{figure}[htbp]
        \vspace{8cm}
        \includegraphics{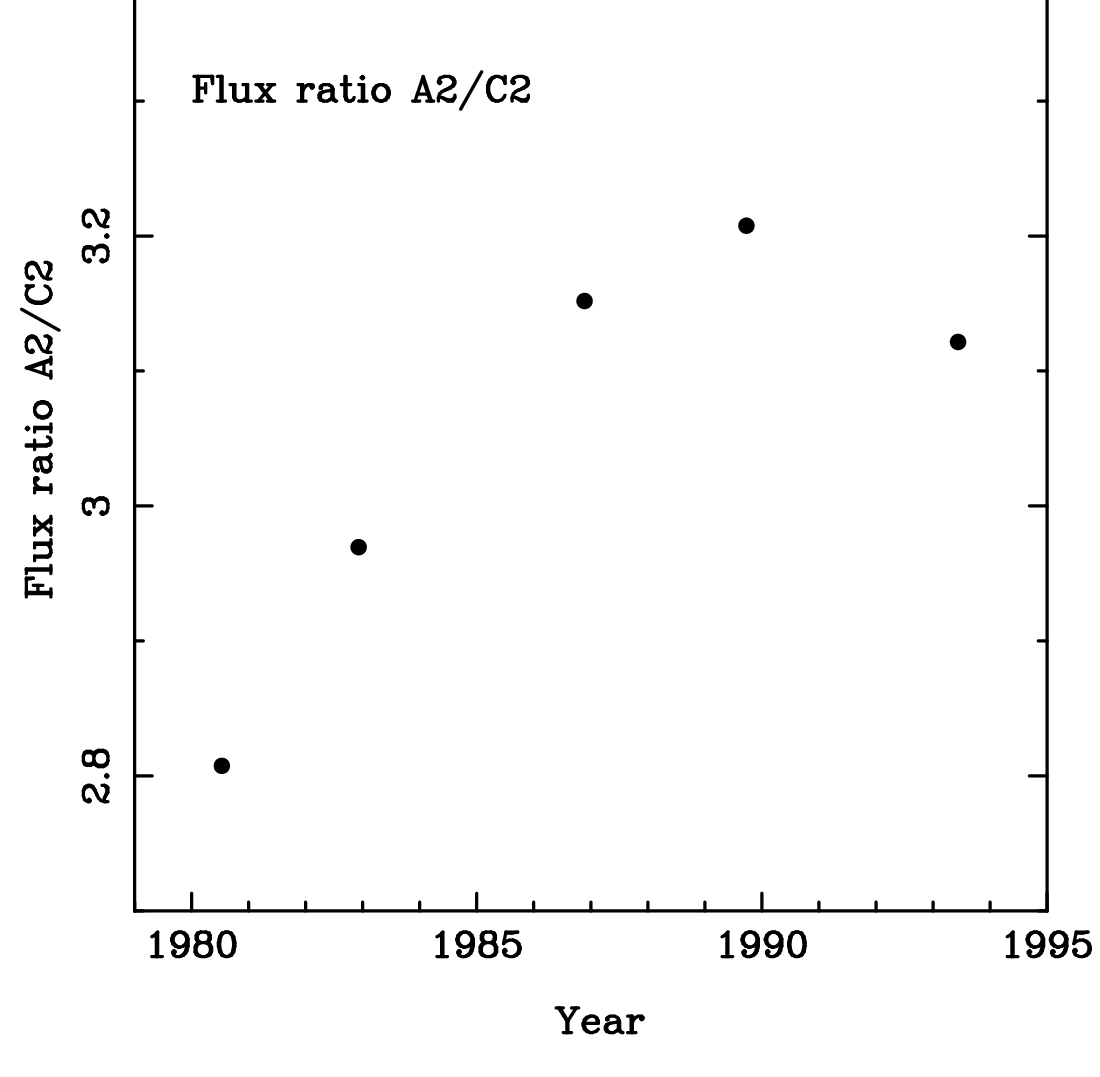}
        \vspace{5mm}
\caption{{\footnotesize Measurements of the flux density ratio
        of A2 and C2 components as a function of time}}
\label{Fig.4}
\end{figure}

One possibility  that must be eliminated is that the apparent change in the
centroid  position  of A2 is  not  due to  motion  of the  whole  component
northward  but instead is due to changes in its internal  structure.  There
could for instance be changes in the relative flux  densities of stationary
subcomponents  within A2.  However  the  non-monotonic  change of the total
flux density of A2 (see  Fig.~\ref{Fig.4})  seems to be inconsistent with a
linear  change of the centroid  position of A2 (see  Fig.~\ref{Fig.3}).  It
also  seems  unlikely  that if  there  were  two  subcomponents  within  A2
separated by a large enough distance to explain the detected centroid shift
that we could  still get a good fit at every  epoch with a single  gaussian
component.  Furthermore we would expect to see changes in the apparent size
of the component  with epoch which we do not see.  Another  possibility  is
that A2 consists  internally  of a true  stationary  hotspot and a jet knot
which moves toward it.  Again in this case we would have difficulty fitting
A2 with a single  gaussian  and would  expect to see the width of the whole
component becoming smaller with time, which we do not see.  While it always
possible to construct  `Christmas Tree' models in which the brightening and
dimming   of   stationary    components   mimics   bulk   motion   (Scheuer
\cite{scheuer84}),  such  models  would  need to be  contrived  to fit  the
observed  changes in  0710+439.  We argue that the changes seen are instead
due to bulk motion of A2.


\subsection{Motions relative to the core}

Our analysis of the  component  motions  showed (see Sect.  4.2) a group of
components  (A1, A3, B2, B4, C1, C2) which are stationary  relative to each
other.  Because we were not able to detect the core position in our maps or
models  (except  tentatively  in the 5th  epoch, see Sect.  3.3), we do not
know for certain the motion, if any, of these  components  relative  to the
core.  Despite  this it seems  most  likely  that this group is  stationary
relative  to the core and  defines  a rest  frame.  If this is not the case
then all these  components, on different  sides of the source, must move in
unison,  in a  coordinated  way  relative  to the core; a  situation  which
appears very unlikely.  Consider for instance if component C2 and hence the
rest of the  stationary  group  moves  southward  from  the  core at  0.126
$h^{-1}c$  (exactly half of the A2-C2 separation  rate).  This scenario has
the advantage  that both A2 and C2 components are then advancing  away from
the core at the same  speed.  But in this case  features B2 and B4 would be
moving  southward  {\it  towards}  the  core at 0.126  $h^{-1}c$.  Since we
believe that B2 and B4 are jet features the probability  that both would be
moving inward toward the core at the same speed, with an amplitude  exactly
matching the advance speed of A2 and C2 outward, would seem to be very low.
The stationary group of components  might  conceivably  move northward, but
then  component  C2 would be moving  inwards to the core, which seems to be
unphysical if C2 is, as we expect, the southern hotspot.

We conclude that the most likely scenario is that all the components in the
`stationary  group' are also stationary with respect to the core as well as
with  respect  to  each  other.  In  this  case  B2  and B4  are  naturally
interpreted  as  stationary  shocks  within  the jet, and B3 as a  possible
travelling  shock  moving  outward  along  the  jet  at  0.1$h^{-1}c$  (see
Sec.4.2).  However a consequence  of this model is that the advance  speeds
of the two outer  components, A2 and C2 through the surrounding  medium are
likely to be different.  Consider the separation  rates of A2-B2 and C2-B2,
which give the lowest estimated errors (see Table~\ref{Tab.3}) of A2 and C2
motion  relative to a  stationary  component.  From these  measurements  we
estimate an advance speed for A2 of $0.230 \pm 0.053 h^{-1}c$ and for C2 of
$0.025  \pm 0.061  h^{-1}c$.  If jet  component  B2 has  after  all a small
undetected velocity outward relative to the core then the implied asymmetry
in the advance speeds of A2 and C2 relative to the core would be increased.
We note however that despite the apparent velocity  difference  statistical
tests can only exclude the null  hypothesis  of no  difference  between the
A2-B2 and C2-B2 advance  speeds at the 10\%  confidence  level; this result
must therefore be confirmed by future observations.

\section{Discussion}

\subsection{Hotspot advance speeds and source age}

The fact that the A2 and C2 gaussian  components are compact and lie at the
leading edges of the minilobe  emission  regions  strongly argues that they
should be  interpreted  as  hotspots  where two  oppositely  directed  jets
terminate  (see  Fig.~\ref{Fig.1}a,b).  In  Sect.  4.2  we  considered  the
relative  separation  rates of the  different  components of 0710+439,  and
found  the  highest  significance  detection  to be  that  of the  apparent
separation  rate of A2-C2 at $0.251  \pm 0.029  h^{-1}c$.  Since CSOs are
dominated  by  unbeamed  emission we expect for a sample  selected on total
flux density that the angle between the jet axis and the sky plane is about
30{\degr}.  Such  an  orientation  for  0710+439  is  compatible  with  the
hotspots  appearing  at the  extreme  ends of the  radio  emission  and not
superimposed  on the  diffuse  minilobe,  and  is  also  consistent  with a
hotspot-core  arm-length  ratio  which  is  close  to one  (Readhead  et al
\cite{read96c}).  Given such an  orientation  we expect that the average of
the two hotspot  advance speeds through the external  medium should be only
slightly larger than that estimated from the observed separation rate (i.e.
approximately $0.126 h^{-1}c$).

Of prime  astrophysical  importance  is the  estimated  age of 0710+439.  A
simple estimate based on dividing the overall projected size of 0710+439 by
the measured  projected  hotspot  separation rate gives $1100 \pm 100$ yrs,
implying  that  0710+439 is a very young radio  source.  In  assessing  the
reliability  of this  estimate  we must be aware  that  what have  actually
measured  (see Sect.  4.2) is the  instantaneous  rate of separation of the
two  hotspots at the present  epoch; not directly the mean rate of increase
of size of the whole source.  Since hotspots are nearly always observed, as
in  0710+439,  to lie  close to the ends of the  lobes in  which  they  are
embedded the mean  expansion  speed of the whole source must equal the mean
separation  rate of the  hotspots.  However  there are  several  mechanisms
which would cause the {\it  instantaneous}  hotspot  advance speeds to vary
about  their  mean  values.  Variations  in the  external  density  due  to
encounters  with clouds is one obvious  mechanism.  Another  possibility is
the so-called  'dentist's drill' phenomena  (Scheuer  \cite{scheuer82})  in
which the position of hotspot's  working surface moves around either due to
jet   precession   initiated   at  the  central   engine  or   hydrodynamic
instabilities  acting  on the  incoming  jet.  In this  model  the  hotspot
executes  a  corkscrew  like-path  in space and the mean rate of advance is
significantly  smaller than the total  instantaneous  speed of the hotspot.
Related  effects  show  up  in  recent   three-dimensional   numerical  jet
simulations  (Norman  1996).  These  simulations  also show  variations  in
hotspot  pressures  and hence  advance  speed due to the  effects of vortex
shedding and cocoon  turbulence  acting on the  incoming  jet which in turn
effect the jet collimation and the area over which the thrust of the jet is
deposited.  Due to such effects the simulations  predict that instantaneous
forward hotspot advance speeds (i.e.  in the direction of the jet axis) can
vary by factors of order two (Norman 1996).

Whatever the physical mechanism that is acting there is empirical  evidence
that in 0710+439  instantaneous  and mean hotspot speeds are different.  In
0710+439 as in other CSOs the  distances  from the two hotspots to the core
are very  similar,  i.e.  the ratio of these  distances  (the  `arm  length
ratio') is only $0.95$ (Readhead et al.  \cite{read96c})  which implies the
mean advance  speeds of the two hotspots have been the same  averaged  over
the history of the source.  In contrast  as  discussed  in Sect 4.4.  it is
probable that the present  instantaneous  advance  speeds for A2 and C2 are
different  from  each  other.  Below  we  discuss  which  of  the  possible
mechanisms is operating in 0710+439 and its impact on our age estimate.

The simplest  explanation for the instantaneous speed variations is that C2
is  presently  interacting  with a dense  cloud  while A2 advances  rapidly
through  an  intercloud  medium.  If around  $100$  such  clouds  have been
encountered  by each  hotspot  there would be enough  cloud  encounters  to
explain  why the arm  length  ratio is close to  unity,  yet few  enough to
explain the  constant  velocity  of A2 over 13 years.  If cloud  collisions
were the reason for hotspot speed variations then our best estimate for the
source age would  depend  inversely  on the  fraction of time, {\it f}, the
hotspots spent transversing the intercloud medium.  We can argue that since
we  apparently  detect  such an advance  in the first CSO for which we have
more than a decade of monitoring {\it f} is unlikely to be less than $0.1$,
and so we obtain an upper age limit of 11\,000 yrs.  Arguing against such a
cloud  mechanism  operating  is the  expectation  that if C2 were  embedded
within a dense  cloud one might  reasonably  expect  its  pressure  to {\it
increase}  in response to the higher  density, in fact we observe  that the
pressure  in C2 is less than in A2.  In addition  we believe it is unlikely
(although  still   possible)  that  the  'dentist's   drill'  effect  is  a
significant  cause of the apparent speed variations  since as we discuss in
Sect 4.1.  the change in the apparent  A2-C2 vector is mainly in its length
and not its  orientation.  Except for certain unlikely  orientations of the
source and its hotspots in space one would expect hotspots  effected by the
dentist drill phenomena to show significant side-to-side motions.

Our favoured  explanation for hotspot speed differences in 0710+439 is that
these are simply  related to  differences  in pressures of the hotspots and
hence  in the  corresponding  ram  pressure  velocities  through  a  smooth
external  medium.  Such pressure  variations and resulting speed variations
are as we  have  noted  predicted  by  recent  three-dimensional  numerical
simulations  (Norman 1996).  Assuming that A2 is close to its equipartition
pressure  (supported  by the analysis of the  frequency of its  Synchrotron
Self Absorbed turnover, Conway et al.  \cite{conway92}) and that the source
is  orientated  not too far from the sky plane then ram pressure  arguments
imply an  external  density  of 1.83  $h^{18/7}$  cm$^{-3}$.  This  value is
similar to that  estimated in the CSO 2352+495 (3 - 10 cm$^{-3}$,  Readhead
et al.  \cite{read96a}) and is consistent with what is expected for the NLR
intercloud  medium.  The  data  are  consistent  with C2  having  the  same
external density as A2 and a lower advance speed simply as a consequence of
its lower pressure.  Since the equipartition pressure of C2 is $0.3$ of A2,
the expected  advance  speed for the same  external  density is 0.55 of A2,
given the A2-C2  separation  rate this implies a C2 advance speed of $0.089
\pm  0.010h^{-1}c$,  which  is  within  $1\sigma$  of  the  observed  B2-C2
separation  rate  of  $0.025  \pm  0.061  h^{-1}c$.  If  hotspot   pressure
variations are the cause of the hotspot speed  variations then the observed
differences in pressure between the two hotspots within  individual CSOs of
between  4 and 6 (see  Readhead  et al  1996b)  imply,  assuming  the  same
external  densities around each hotspot, that hotspot speed variations vary
over a factor of about two within each source.  We therefore  expect ratios
between instantaneous and mean separation rates to be of the same order and
hence  estimate  an upper  limit to the age of  0710+439  of  approximately
$3000$yrs.

Given our age  estimates  and estimates of the jet thrust  (Readhead et al.
\cite{read96a}) we can compare the mechanical  luminosity required to drive
the hotspots forward with the radio  luminosity and jet power.  For an age
of  1100yrs  the  combined  mechanical  luminosity  of the two hotspots is
$0.8\times10^{44}  h^{-17/7} $ erg s$^{-1}$, while the radio  luminosity of
the  two  hotspots  is  about   $0.5\times10^{44}   h^{-2}$  erg  s$^{-1}$.
Following the arguments used in Readhead et al.  (\cite{read96a})  from the
measured  hotspot  sizes and  pressures  the upper limit on the total power
supplied by the jets is $8.0\times10^{44}h^{-10/7}$  erg s$^{-1}$.  A lower
limit on the total jet power can be obtained by adding  together  the radio
power  and  mechanical  work.  The  total  jet  luminosity  is (for  h=0.6)
therefore    in    the    range    $4.5\times10^{44}$erg     s$^{-1}$    to
$16.6\times10^{44}$erg  s$^{-1}$ and the  efficiency  of  conversion of jet
energy  to  radio  emission  is  between  8\% and  31\%.  In  contrast  for
classical  FRII  (Fanaroff  \& Riley  \cite{fanaroff})  radio  galaxies  we
estimate upper limits on hotspot radiative efficiencies of a few percent by
comparing  total radio  luminosities  to estimates of the jet  luminosities
given by Rawlings \& Sanders (\cite{rawlings}).

\subsection{Implications for CSO models}

Our best estimate for the mean hotspot advance speed in 0710+439, given our
observations, i.e  $0.13h^{-1}c$, is somewhat larger than that estimated by
other  authors  for the CSO  population  in general  (e.g.  Readhead et al.
\cite{read96b}  estimates $0.02c$).  If hotspot pressures and hence advance
speeds  vary with  time it might be that the true  mean  advance  speeds in
0710+439 are up to a factor of two less than our best  estimate  (see Sect.
5.1) but a difference  between  predictions and observations still remains.
One possibility, given that one would expect a range in properties from CSO
to CSO, is that 0710+439 lies at the extreme end of the  population  and is
growing   faster   than   the   typical   CSO.  However,   Conway   et  al.
(\cite{conway94})  tentatively  detected,  based on two global 5 GHz epochs,
mean hotspot advance  velocities of $0.09h^{-1}c$ in another CSO, 0108+388.
A similar  rate of advance was also  detected by Taylor et al (1996) in the
same    source.   Recently    a   mean    hotspot    advance    speed    of
$0.098\pm0.013h^{-1}c$ has been confirmed in 0108+388 by three epoch global
5 GHz  observations   (Owsianik  et  al.  \cite{owsianik}).  Conway  et  al.
(\cite{conway94})  also detected a hotspot advance speed of  0.065$h^{-1}c$
in the  object  2021+614  which  may  also  be a CSO.  Finally  for the CSO
2352+495  Readhead et al.  (\cite{read96a})  gives age  estimates of 1200 -
1800 yrs based on synchrotron ageing and 1500 - 7500 yrs from energy supply
arguments.  For this source of size 120$h^{-1}$pc an age near the lower end
of the allowed  range, i.e.  $1500$ yrs gives a mean hotspot  advance speed
of $0.13h^{-1}c$.

The lower  estimate of hotspot  advance  speeds for the CSO  population  in
general ($0.02c$) made by Readhead et al.  (\cite{read96a})  was based on a
two  part  argument,  namely:  i) it  was  argued  that  hotspot  pressures
adjusted  to the  external  density  so that  hotspot  advance  speeds  are
constant.  Therefore  advance  speeds of high  pressure  hotspots  in young
sources  transversing the dense ISM are the same as in the classical double
sources; ii) Classical  double sources, based primarily on observations  of
Cygnus A, have advance speeds of $0.02c$.

The first part of the above argument was based on detailed  observations of
three CSOs, in which the arm-length  ratios are close to one and therefore
the mean advance  speeds for the two hotspots must be the same,  despite in
each  case  the  pressures  of the  two  hotspots  being  quite  different.
Readhead   et   al.   (\cite{read96b})    explicitly   assumed   that   the
characteristics  of the hotspots are constant in time and that the pressure
ratios  measured now are typical of the whole history of these sources.  It
follows  that  hotspot  advance  speeds  must  be  independent  of  hotspot
pressure.  It was  postulated  that this could be achieved  if a  mechanism
existed where the hotspot pressure always adjusted to the external  density
so that ram pressure balance gave a constant advance speed.

In contrast 3-D numerical simulations (Norman \cite{norman})  indicate that
due to  hydrodynamic  effects  individual  hotspots can rapidly  vary their
pressures  around some mean value as they move outward, with  corresponding
variations in their ram-pressure  advance speeds.  Differences in pressures
between  hotspots seen in maps may therefore be just temporary  features of
sources.  Arm length  ratios close to one are simply  explained if external
densities  and mean  hotspot  pressures  on each side of the source are the
same, so that mean advance speeds are the same.  It follows that no special
mechanism is required which adjusts hotspot pressure to external density in
order to explain the  observations.  The main motivation which led Readhead
et al.  (\cite{read96b})  to propose a universal  constant  hotspot advance
speed for both CSOs and classical sources is therefore removed.

In contrast to Readhead et al.'s  (\cite{read96b})  observational  approach
Begelman  (\cite{begelman})  has  calculated  the evolution  expected for a
simple theoretical model of a source with an over-pressurised  cocoon and a
hotspot  whose mean  pressure is a fixed  ratio to that of the  cocoon.  In
this model the advance speed depends on the density versus  distance of the
external  medium $\rho \propto  r^{-n}$, such that the advance speed $v_{h}
\propto l^{\beta}$ where $l$ is the source size and $\beta = (n-2)/3$.  For
n in the  plausible  range  $1.5$ to $2.0$,  then  $\beta$  is in the range
$-0.17$ to $0.0$.  It is therefore  possible that hotspot advance speeds in
CSOs are  somewhat  faster than in classical  sources.  Since CSOs are 1000
times  smaller  than  classical  sources  if n were 1.5, we expect  advance
speeds which are about 3 times faster.

Readhead et al.  (\cite{read96a})  estimated  advance  speeds in  classical
sources  to be  $0.02c$,  mainly  based on  Cygnus A  results.  However  it
appears that Cygnus A is an unusual  source in that it lies in an unusually
dense  environment  (Barthel \& Arnaud  \cite{barthel},  Reynolds \& Fabian
\cite{reynolds}).  In other FRII's  external  densities are estimated to be
$30$ times  smaller  (Rawlings  \&  Saunders  \cite{rawlings})  yet hotspot
pressures  are only $3$ times  smaller  (Readhead  et al.  \cite{read96b}),
implying  that typical ram pressure  advance  speeds in  classical  sources
might be closer to $0.06c$.  Hotspot  advance  speeds can also be estimated
independently  from electron  spectral ageing arguments and from arm-length
asymmetries in classical  double  sources.  Using the first method the data
of Rawlings \& Saunders (\cite{rawlings}) indicate advance speeds of $0.108
\pm 0.098 h^{-4/7}c$;  other studies indicate  velocities which are greater
than $0.1c$ (e.g.  Liu et al.  \cite{liu}).  Such  estimates  might however
be larger then the real advance speeds since strictly speaking they measure
the sum of the advance  speed of the hotspot and the speed of the  backflow
from it (see Liu et.  al.  \cite{liu} and Scheuer  \cite{scheuer95}).  This
would be  consistent  with the fact  that for the same  sample  of  sources
observations of  jet/counter-jet  side arm length ratios indicate  (Scheuer
\cite{scheuer95})  smaller  advance speeds of  $0.03\pm0.02c$.  The present
data on hotspot  advance speeds does not yet yield a define  conclusion but
certainly  allows the  possibility  that these speeds  could be a factor of
three larger than estimated by Readhead et al.  (\cite{read96a}).

Combining a typical FRII  advance  speed of say $0.06c$  with the  probable
weak evolution of hotspot advance speeds with source size we find that mean
hotspot  advance  speeds in CSOs can  plausibly  be $0.2c$ or  larger.  We
conclude  that  the size of the  measured  hotspot  speed  in  0710+439  is
compatible  with the  predictions of theoretical  models.  Such fast speeds
imply that sources  have only a short  lifetime in the CSO phase.  The fact
that up to 10\% of  sources  in flux  limited  samples  at 5 GHz  are  CSOs
therefore means either that i) not all CSOs evolve into classical  sources;
some  exhaust  their fuel  before  reaching  100kpc size  (Readhead  et al.
\cite{read94})  or ii) there is strong luminosity  evolution in their radio
emission.  We favour the second explanation, strong luminosity evolution of
the  required  amount to explain the source  size  distribution  is in fact
predicted  by  the   theoretical   models.  For   instance   the   Begelman
(\cite{begelman})  model  predicts  a  radio  luminosity   proportional  to
approximately $l^{-0.5}$ assuming a constant jet mechanical power.  For the
weakly evolving hotspot advance velocity  predicted for an external density
of the form $\rho \propto  r^{-1.5}$,  the  predicted  number of sources in
each  decade  of size  then  exactly  matches  the  observations  (Begelman
\cite{begelman}).  As first noted by Readhead et al.  (\cite{read96a})  for
2352+495,  and as we find for 0710+439  (see Sect.  5.1), the limits on the
radiative  efficiency  for CSOs compared to classical  sources  empirically
demonstrate that the expected  luminosity  evolution does in fact occur and
with a magnitude (a factor of 30 from CSO to classical sources)  consistent
with that  expected by theory.  Given this  efficiency  evolution one would
expect   0710+439   to   evolve   into  a  source   of   radio   luminosity
L$_{178MHz}=8\times10^{25} h^{-2}$  W  Hz$^{-1}$,  i.e.  a weak FRII. 
 We conclude
that CSOs are  probably  very young  extragalactic  radio  sources and that
furthermore  they  probably  evolve into lower  luminosity  FRII  classical
double radio sources.

\begin{acknowledgements}
     
We thank the  observatories  of the US and European VLBI Networks and NRAO,
which  operates  the VLA  and the  VLBA.  NRAO is  operated  by  Associated
Universities,  Inc., under cooperative  agreement with the National Science
Foundation.  I.  Owsianik   acknowledges  support  from  the  Polish  State
Committee  for  Scientific   Research  grant  nr  2.P304.003.07,  EU  grant
(ERBCIPDCT940087)  and Onsala Space  Observatory.  We thank the referee for
his helpful  comments.  We thank G.B.  Taylor for helping schedule the 1993
observations.

\end{acknowledgements}

\appendix

\section{Appendix}

We must be careful with the  interpretation  of  multi-epoch  data, because
analysis methods can strongly affect the reliability of our results.  It is
very  important to know which  features to believe,  and how  accurate  our
measurements  of very small changes are.  To investigate  these problems we
made mapping and  modelfitting  tests in which we generated  simulated data
using the program FAKE in the Caltech VLBI data  analysis  package with the
same  {\it u-v}  coverages  as the real  observations.  These  `fake'  data
contained  realistic  additive Gaussian noise, random  antenna-based  phase
errors and time-varying amplitude calibration errors of order of 10\%.

\subsection{Reliability of features in CLEAN maps}

Fake data for the 1st and 5th epoch  were made  using  the same 9  gaussian
components  model,  which  was  fitted  to the real  5th  epoch  data  (see
Table~\ref{Tab.2}).  These simulated data were each mapped  separately with
a point  component  as a  starting  model.  The  final  maps had  different
extended  structures  in the northern  and southern  components,  they also
contained  apparent bridge  emission  between these three main  components,
which was not part of the model.  These detected errors were presumably due
to  inadequacies  and  differences  in {\it  u-v}  coverage  at each  epoch
combined  with  differences  in the details of mapping at each `fake' epoch
(choice of windows  etc).  This test  leads us to the  conclusion  that the
details of diffuse and bridge  structures  seen in the CLEAN maps made from
the real data (see  Fig.~\ref{Fig.1}a)  are not reliable.  We also conclude
that  using  maps made  separately  at each epoch is an  unreliable  way of
detecting changes in source structure.

\subsection{Estimating modelfitting component position errors}

As described in Sect.  4.1 we can set accurate limits on component  motions
by gaussian  modelfitting to each epoch.  We carried out `fake' simulations
to answer two questions  about this  procedure.  The first was to determine
the  size  of the  remaining  random  errors  due to  different  {\it  u-v}
coverages and data reduction at each epoch under the null  hypothesis of no
component  motion.  The second  question was to investigate  if this method
introduced  systematic errors by biasing the results in the sense that real
changes  would be removed or reduced by initially  trying to force the data
at each epoch to agree with the same 5th epoch starting model.

In our first test in order to make the  simulation as realistic as possible
we attempted to take account of the fact that the real source  structure in
0710+439 is almost  certainly  more  complex than can be  represented  by 9
Gaussian components.  This complexity is demonstrated by the fact the final
agreement  factors  (see Sect.  4.1) of our models are  further  from unity
than would be expected  purely from random noise.  It is  conceivable  that
this extra complexity might interact with differences in {\it u-v} coverage
to give apparent changes in the centroid  position when a 9 component model
is fitted at each epoch, even if no position  changes actually occur in the
source.  To test the above  possibility we created a 18 component  model by
replacing each component in the original 9 component  model by two slightly
shifted  gaussians.  The final model was such that the agreement factors on
fitting a $9$ component model to the corresponding  `fake' data was roughly
similar to that obtained with the real data.

Having chosen a suitable 18 component  model we created `fake' data for the
1st and 5th epochs.  Using a procedure  as similar as possible to that used
to analyse  the real data we then fitted a 9 component  model at each epoch
and compared the  separations  between the gaussians we obtained.  The size
of the apparent  changes gave us an estimate of the residual  random error.
On doing this test we found an apparent change of A2-C2  separation of 8.09
$\mu$as and in the A2-B2 separation of 30.9 $\mu$as between the 1st and the
5th  epochs.  The values are much  smaller  than the  changes in  component
separation which we detected from the real data (see  Fig.~\ref{Fig.3}) and
comparable with the variances we estimated from linear regression  analysis
(see Sect.  4.2).

\subsection{Biasing  due to cross  self-calibration}  

In our final  test we sought to  determine  if our  modelfitting  procedure
introduced a systematic error due to initially  self-calibrating all models
against  the same 5th  epoch  starting  model.  It is  possible  that  real
changes might be reduced or removed by initially trying to force all epochs
to agree with the 5th epoch  model.  To quantify  this effect we  simulated
the case of a 200 $\mu$as shift of A2 (and then C2) position between epochs
1 and 5.  Applying our standard modelfitting  procedure (see Sect.  4.1) we
determined  an  estimated  motion of just under 200  $\mu$as.  The  largest
negative bias found in our test was only 9 $\mu$as.  We conclude  that this
biasing  mechanism has a negligible  effect on the estimate in the shift of
A2 seen in the real data.

\end{document}